\documentclass[preprint]{aastex}
\usepackage{epsfig}
\usepackage{rotating}

\shorttitle{On the radio spectra of supernova remnants}
\shortauthors{D. Uro{\v s}evi{\' c}}

\begin{document}

%% LaTeX will automatically break titles if they run longer than
%% one line. However, you may use \\ to force a line break if
%% you desire.

\title{ON THE DETERMINATION OF THE EVOLUTIONARY STATUS OF SUPERNOVA REMNANTS
FROM RADIO OBSERVATION DATA}

%% Use \author, \affil, and the \and command to format
%% author and affiliation information.
%% Note that \email has replaced the old \authoremail command
%% from AASTeX v4.0. You can use \email to mark an email address
%% anywhere in the paper, not just in the front matter.
%% As in the title, use \\ to force line breaks.

\author{Dejan Uro{\v s}evi{\'c}}

\affil{Department of Astronomy, Faculty of Mathematics,
University of Belgrade, Studentski trg 16, 11000 Belgrade, Serbia}
\email{dejanu@math.rs}

%\affil{$^{2}$Isaac Newton Institute of Chile, Yugoslavia Branch}

%% Notice that each of these authors has alternate affiliations, which
%% are identified by the \altaffilmark after each name.  Specify alternate
%% affiliation information with \altaffiltext, with one command per each
%% affiliation.

%% Mark off your abstract in the ``abstract'' environment. In the manuscript
%% style, abstract will output a Received/Accepted line after the
%% title and affiliation information. No date will appear since the author
%% does not have this information. The dates will be filled in by the
%% editorial office after submission.

\begin{abstract}

This paper aims to give a brief review of a new concept for the preliminary determination of the evolutionary status of supernova remnants (SNRs). Data obtained by radio observations in continuum are used. There are three different methods underlying the new concept: the first one based on the location of observationally obtained radio surface brightness and corresponding diameter of an SNR on the theoretically derived $\Sigma-D$ tracks; the second one based on the forms of radio spectra; and the third one, based on the magnetic field strengths that are estimated through  the equipartition (eqp) calculation. Using a combination of these methods, developed over the last two decades by the Belgrade SNR Research Group, we can estimate the evolutionary status of SNRs. This concept helps radio observers to determine preliminarily the stage of the evolution  of an SNR observed in radio domain. Additionally, this concept was applied for several SNRs, observed by the Australia Telescope Compact Array (ATCA), and the corresponding results are reviewed here. Moreover, some of the results are revised in this review to reflect the updated recently published $\Sigma-D$ and eqp analyses.

% \PACS{PACS code1 \and PACS code2 \and more}
% \subclass{MSC code1 \and MSC code2 \and more}
\end{abstract}


\section{Introduction}
\label{intro}

Radio continuum observations of supernova remnants (SNRs) help us obtain directly  their flux densities and angular extensions. From these two quantities we can easily calculate the so-called surface brightness $\Sigma$, which is a distance independent quantity. The surface brightness of an SNR changes over time. On the other hand, with time propagation, the volume of an SNR (defined by approximately spherical shock wave which separates the object and surrounding interstellar medium) increases. If we assume an SNR is of a spherical form, we can define its diameter $D$. As a result, the evolution of an SNR can be described by changes of the surface brightness with the increase of its diameter i.e. by the so-called $\Sigma-D$ relation. The method based on radio surface brightness evolution of an SNR with the increase of its diameter is the first ingredient of the concept, introduced in this review, for determining the evolutionary status of an SNR. The second one relates to the different forms of SNR radio continuum spectra. The radio continuum spectra of SNRs are derived from the observations in different frequency bands - they are defined by the values of flux densities at different frequencies. A spectrum is better if we can provide more observations at different frequencies. The forms of SNR radio spectra are different in different stages of evolution of an SNR. Due to this we can estimate the evolutionary status of an SNR from a careful analysis of its spectrum. Finally, the third method is connected to the equipartition (eqp) calculation   magnetic field strengths in SNRs - namely, younger SNRs  have conditions for  higher magnetic fields, while older SNRs have lower magnetic fields. Each of these three methods is based on the diffusive shock acceleration (DSA) theory. This type of particle acceleration is responsible for the production of cosmic rays (CRs) at the SNRs strong shock waves. All of these three methods are presented in detail in the next three sections of this review, respectively. In Section 5 the concept of a combined use of these three methods for determining the evolutionary status of SNRs is given. Examples from published papers (the first one in 2012) in which this new concept was applied are presented in Section 6. Pavlovi{\' c} et al. (2018) and Uro{\v s}evi{\' c} et al. (2018) re-analyzed the $\Sigma-D$ and eqp methods and due to these new results, the estimated evolutionary statuses from Section 6 (eight SNRs; papers published from 2012 to 2018) are revised in Section 7. Section 8 summarizes the previously presented. The short version of this study was presented in Uro{\v s}evi{\'c} (2020), where only the fundamental ideas of the new concept were given.

\section{$\Sigma-D$ tracks}

\label{sec:1}
%Text with citations \cite{RefB} and \cite{RefJ}.

The $\Sigma-D$ relation for SNRs was defined by Shklovsky (1960a,b) - in two directions, as a method for description of the SNR radio surface brightness evolution, and as a method for determination of distances to SNRs. This relation was studied for almost sixty years, with more or less activity in individual decades. The first four decades of development of $\Sigma-D$ studies are reviewed in Uro{\v s}evi{\' c} (2000, 2002, 2005). In the last two decades Uro{\v s}evi{\' c} et al. (2003a,b), Berezhko \& V\"olk (2004, hereafter BV04), Uro{\v s}evi{\'c} \& Pannuti (2005), Bandiera \& Petruk (2010), Pavlovi{\' c} et al. (2013, 2018) developed theoretical concepts of $\Sigma-D$ studies. The authors who worked on the development of empirical relations in the last two decades include: Guseinov et al. (2003), Uro{\v s}evi{\'c} (2003), Arbutina et al. (2004), Arbutina \& Uro{\v s}evi{\' c} (2005), Uro{\v s}evi{\' c} et al. (2005, 2010), Bandiera \& Petruk (2010), Pavlovi{\' c} et al. (2013, 2014), Vukoti{\' c} et al. (2014), Kosti{\'c} et al. (2016), Bozzetto et al. (2017), Vukoti{\' c} et al. (2019).

As mentioned in the introduction, the radio surface brightness is a quantity independent of the distance to the object. Here we encounter the first problem in this method, because to calculate the diameter, we also need distance to the object. For extragalactic samples of SNRs, this problem does not exist - all SNRs in one galaxy are at the same distance equal to the distance between us and that galaxy. This problem, however, does exist for the Galactic SNR sample. We do not have precise methods for estimating SNR distances in our Galaxy. Due to this, the diameter of an SNR can be estimated only with significant uncertainty. There are several independent methods (not based on the $\Sigma-D$ relation) for determining distances to the Galactic SNRs. We can use the SNRs with independently determined distances for the creation of a calibration sample. For the calibration sample, we can set $\Sigma-D$ relation and use this relation to estimate the distances to SNRs for which we do not have independently determined distances. Around 100 SNRs in the Galactic sample have independently determined distances, but for approximately 200 of them we have to use $\Sigma-D$ relation  in order to determine their distances (for details see Vukoti{\' c} et al. 2019).

 In this review we examine how the radio surface brightness of an expanding SNR evolves, i.e. the corresponding $\Sigma-D$ dependence. The main idea for the method of determination of the evolutionary status of an SNR by using $\Sigma-D$ tracks is based on theoretically derived changes of radio surface brightness of an expanding SNR (forming of evolutionary paths). For initial conditions, we can set the values for explosion energies of supernovae and values for densities of surrounding media in which SNRs expand. For different combinations of energies and densities we can obtain different evolutionary paths. For extragalactic SNRs, the radio surface brightness and diameter are obtained directly from observations - diameters can be easily calculated if we know distance to the host galaxy. If we analyze a Galactic SNR, the distance should be estimated first (by using some independent method or $\Sigma-D$ relation) and then we should calculate the diameter. An observed SNR with the determined values for $\Sigma$ and $D$, can be shown in the $\Sigma-D$ plane, and after examining on which evolutionary path this SNR is located, we can estimate its evolutionary status, environmental density, and SN explosion energy.

BV04 in their theory of synchrotron emission from supernova remnants, for the first time presented the evolutionary paths in the $\Sigma-D$ plane. This analysis is based on the time-dependent nonlinear kinetic theory for particle acceleration in SNRs. They used numerical calculations performed for the expected range of ambient densities and supernova explosion energies. The magnetic field in SNRs is assumed to be significantly amplified by nonlinear DSA effects.

In the next and the last analysis of the radio evolution of SNRs based on nonlinear kinetic theory of CR acceleration coupled with three-dimensional hydrodynamic simulations, Pavlovi{\' c} et al. (2018, hereafter P18) took a new approach in the creation of evolutionary $\Sigma-D$ tracks. They performed simulations for a wide range of the relevant physical parameters, such as ambient density, supernova (SN) explosion energy, acceleration efficiency, and magnetic field amplification (MFA) efficiency. A detailed description of the supercomputer simulations applied for the creation of $\Sigma-D$ tracks is given in P18.

The $\Sigma-D$ diagrams presented in the previous two papers can be used for determination of the evolutionary status of an observed SNR. The evolutionary paths in P18 are obtained by using advanced approach and due to this they are better for further use.

\section{Forms of radio spectra}

The second method which can be used for determining the evolutionary status of SNRs is based on the analysis of the forms of SNR continuum radio spectra. The details on the SNR radio spectral forms can be found in Uro{\v s}evi{\' c} (2014). Here, only a brief review is presented. Young SNRs have steeper spectral indices with $\alpha > 0.5$ ($S_\nu\propto\nu^{-\alpha}$, where $S_\nu$ is the flux density at the observed frequency $\nu$). These steeper spectral indices are the results of the nonlinear particle acceleration effects with incorporating strong MFA (for details see Uro{\v s}evi{\' c} (2014), Pavlovi{\' c} 2017, Bell et al. 2019). Also, the radio spectra of young SNRs can be curved (concave-up) - again it is the effect of the nonlinear DSA, i.e. modification of the shock front (for details see Uro{\v s}evi{\' c} 2014, and Oni{\' c} \& Uro{\v s}evi{\' c} 2015). Also, the steep linear or curved spectra frequently appear for the evolved SNRs. The test particle DSA predicts linear spectra for older SNRs with spectral indices around 0.5 and they should be steeper with further SNR evolution. Mostly, the evolved SNRs have spectral indices in the interval $0.5 < \alpha < 0.6$. Additionally, the curved, concave-up spectra can be expected for evolved SNRs where a significant amount of thermal bremsstrahlung radiation can be added to the synchrotron radiation. The evolved SNRs in some cases show concave-down spectra. This kind of spectrum can be explained using DSA theory with the effect of synchrotron losses within the finite emission region. Also, for details on the spectra of evolved SNRs see Uro{\v s}evi{\' c} (2014).

Forms of radio spectra can be used for determining the evolutionary status of an SNR, i.e. determining whether it is young or evolved. For easier reference see  Table 1 in Uro{\v s}evi{\' c} (2014).

\begin{table}
% table caption is above the table
\caption{Revised Table 1 from Uro{\v s}evi{\' c} (2014). The theoretically predicted radio spectra and some examples for observational spectra of shell, composite and mixed-morphology SNRs.}
\smallskip
%\label{tab:1}       % Give a unique label
% For LaTeX tables use
\resizebox{\textwidth}{!}{
\begin{tabular}{l|l|l|l|l|l}
\hline\noalign{\smallskip}
\multicolumn{6}{c}{theoretical predictions} \\
\hline%\noalign{\smallskip}\hline\noalign{\smallskip}
\hline%\noalign{}
&\multicolumn{2}{l}{linear radio spectra}&&\multicolumn{2}{l}{curved radio spectra}  \\ \hline
&$\alpha=0.5$&steep ($\alpha>0.5$)&flat ($\alpha<0.5$)&concave-up&concave-down \\
\hline\hline
young SNRs  & test particle & ampl. mag. field + &DSA + &non-linear&obs. effects +\\
&DSA& quasi-perp. shocks&Fermi 2&DSA&DSA effects\\ \hline
evolved SNRs & DSA  & test particle &DSA +&synch. + brem.&obs. effects + \\
& & DSA &Fermi 2&or spin. dust&DSA effects\\
\hline\hline\noalign{\smallskip}
\multicolumn{6}{c}{from observations} \\
\hline%\noalign{\smallskip}\hline\noalign{\smallskip}
\hline%\noalign{}
&\multicolumn{2}{l}{linear radio spectra}&&\multicolumn{2}{l}{curved radio spectra}  \\ \hline
&$\alpha=0.5$&steep ($\alpha>0.5$)&flat ($\alpha<0.5$)&concave-up&concave-down \\
\hline\hline
young SNRs  & / & e.g. Cas A,&/&e.g. Cas A, Tycho, &/ \\
&&G1.9+0.3&&Kepler, SN1006&\\ \hline
evolved SNRs &e.g. Monoceros and &e.g. HB3, HB9 &e.g. W28,&e.g. IC443, &e.g. S147, HB21,  \\
&Lupus loops & &Kes67, 3C434.1&3C391, 3C396&J0455-6838\\
\hline\noalign{\smallskip}\hline

\end{tabular}
}
\end{table}

\section{The magnetic field strengths}

Relying on data obtained by radio observations, with additionally provided distance to an SNR, and using the eqp calculation we can estimate the magnetic field strengths. This method was suggested by Pacholczyk (1970). The modification of the original method was given by Beck \& Krause (2005). Development of the eqp method applied to SNRs was started by Vukoti{\' c} et al. (2007) and continued by Arbutina et al. (2012, 2013) and Uro{\v s}evi{\' c} et al. (2018). The eqp method is based on the calculation of the value of magnetic field strength for which total energy density in the system is minimal (a part of the total energy density is necessary for synchrotron emission). This total energy density has two ingredients: the energy density of CRs and the energy density of magnetic fields. The minimal energy requirement is equivalent to eqp assumption - it means that energy densities of CRs and magnetic fields are approximately equal (eqp). Additionally these two energy densities can be in any constant partition (Arbutina et al. 2012, Uro{\v s}evi{\' c} et al. 2018) and used for determining magnetic field strengths. In a recent paper from this series, Uro{\v s}evi{\' c} et al. (2018) showed  that the eqp (or the constant partition) between ultrarelativistic electrons and magnetic fields represents the best starting assumption for estimating the magnetic field strengths in SNRs. This is shown by using 3D hydrodynamic supercomputer simulations, coupled with a nonlinear diffusive shock acceleration model. However, the eqp method can be used only for the estimation of magnetic field strengths with the order of magnitude precision. For our purposes - determining the evolutionary status of an observed SNR, the order of magnitude determination is precise enough. In young SNRs, where shocks are very strong, nonlinear DSA effects can provide conditions for MFA. The magnetic field amplification in modified shocks can be 100 times higher than by the compression obtained fields in the non-modified shocks. The magnetic fields made only by shock compression can be at most 4 times higher than interstellar (IS) magnetic fields (average value of IS magnetic field is around 5 $\mu$G). Due to this, the order of magnitude of a few tens  of $\mu$G should correspond to the magnetic fields of evolved SNRs. For young SNRs, the characteristic values are of few to several hundred $\mu$G. All of these estimates depend on the density of environment. Discussion on the dependence between density of environment and magnetic field strength is presented in Sections 6 and 7.

The calculation of magnetic field strengths from the eqp model is a rather straightforward process. An observer should provide data obtained from observations: the radio flux density at a radio frequency for an SNR, spectral index, distance, and volume filling factor (part of the volume of an SNR from which we see synchrotron emission - the volume of shell). User should enter data on the web page: \url{http://poincare.matf.bg.ac.rs/$\sim$arbo/eqp/}, and the calculator\footnote{Use $\kappa=0$ for electron eqp, or $\kappa\neq0$ for CR eqp.} will return eqp magnetic field strength and the minimal total energy. The formulae implemented in this calculator and the corresponding detailed explanations can be seen in three papers: Arbutina et al. (2012, 2013) and Uro{\v s}evi{\' c} et al. (2018).

\section{Determination of evolutionary status - a new concept}

To determine the evolutionary status of an observed SNR in radio spectrum, we can start with $\Sigma-D$ analysis. The theoretically derived evolutionary paths from P18 are shown in Figure 1. In this Figure, different line colors correspond to different ambient densities, $n_H$/cm$^3$ = 0.005
(cyan), 0.02 (blue), 0.2 (green), 0.5 (red), and 2.0 (black). Dotted, dashed and solid lines correspond to different explosion energies, namely $E_0$/10$^{51}$ erg
= 0.5 (dotted), 1.0 (dashed), and 2.0 (solid). Observational  data marked with triangles represent 65
Galactic SNRs with known distances  taken from Pavlovi{\' c} et al.
(2014).  They represent the evolutionary tracks for
injection parameter $\xi = 3.4$ and nonlinear magnetic
field damping parameter $\zeta = 0.5$. For more details see P18.
SNR Cassiopeia A (Cas A) is marked with an open triangle, while the youngest Galactic SNR, G1.9+0.3 (see Pavlovi{\' c} 2017 for detailed
modeling) is marked by an open circle.
 Numbers represent the following SNRs: (1) CTB 37A, (2) Kes 97,
(3) CTB 37B, and (4) G65.1+0.6.

\begin{figure}[h]
  \includegraphics{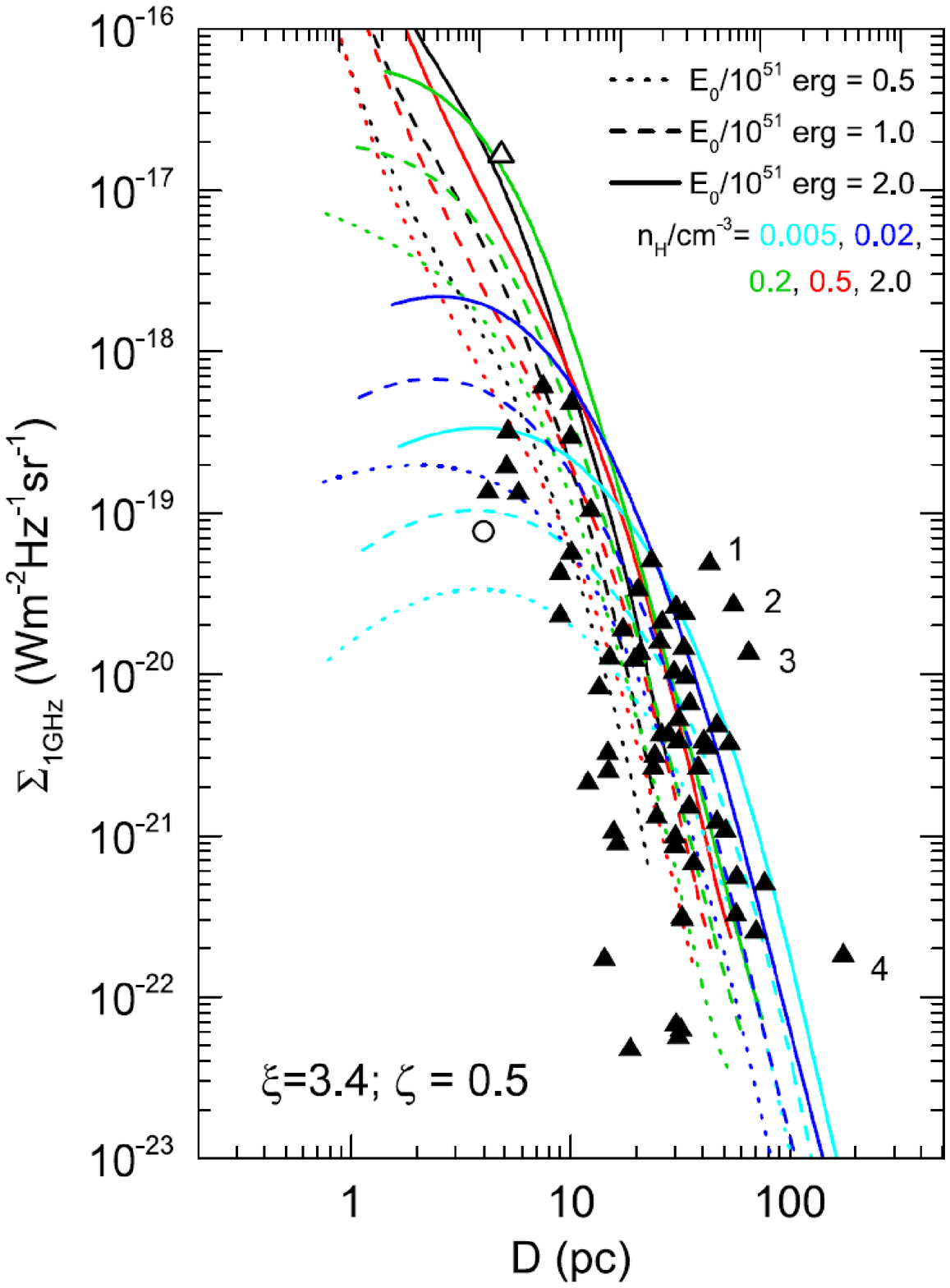}
  \centering
\caption{ $\Sigma - D$ diagram for SNRs at
1 GHz from P18. }
\end{figure}

 A line in Figure 1 represents an evolutionary path for one combination of the SN explosion energy and the ambient density. The ascending part of the evolutionary track  represents the evolution of a very young SNR in the early free expansion phase. When the line starts to decline, this corresponds to late free expansion and after that the SNR enters the early Sedov phase of evolution. When steepness of the evolutionary path becomes the highest and has a constant slope, this part corresponds to the Sedov phase of evolution. The lines finish at the end of the late Sedov phase. The simulations presented by these $\Sigma-D$ tracks do not cover the radiative phases of evolution.

If radio observers detect new SNR in radio and calculate the corresponding surface brightness and diameter, they can use Figure 1 and locate their SNR somewhere in the $\Sigma-D$ plane. Depending on which part of the curved evolutionary path targeted SNR is located in, observers can determine the phase of evolution for their SNR. They can also estimate the relevant ambient density and SN explosion energy.

On the other hand, the evolutionary paths in Figure 1 are very close and intercept each other. Due to this, we are not sure if we have a unique $\Sigma-D$ track for an SNR. We therefore move to next method and check the form of spectrum of a newly observed SNR. As mentioned in Section 3, we should check spectral index value, or whether or not the spectrum is curved, and by using Table 1 and analysis in Uro{\v s}evi{\' c} (2014) we can obtain more information on the evolutionary status of an SNR - whether the SNR is young or evolved. This is one method more that can be combined with the $\Sigma-D$ track method.

Additionally, concave-up spectra can represent young, but they can also represent evolved SNRs (see Section 3 and Table 1). Also, spectral index slopes steeper than 0.5 are a consequence of the strong non-linear effects at the start of SNR evolution, but for SNRs that are evolved steeper spectral slopes may be due to a low efficiency of particle acceleration. If a mix of the $\Sigma-D$ and spectral form methods does not lead to a single conclusion, we should resort to the eqp calculation in order to establish magnetic field strengths. As explained in Section 4, observers should prepare data based on which they can very easily calculate the values of the SNR magnetic fields. The higher the magnetic field, the younger the newly observed SNR. Overall, we can make a preliminary determination of the evolutionary status of an SNR with optimal reliability if we combine the three methods. This combination of methods represents the new concept introduced in this review.

To check how this concept works, we will look at the youngest two Galactic SNRs: Cas A (approximately 330 years old), and G1.9+0.3 (120 years old). Both SNRs were studied many times and we know in which phases of evolution they are.

In Figure 1, Cas A is shown by open triangle, and G1.9+0.3 is shown by circle. We can estimate the evolutionary status of Cas A by using Figure 1: young SNR, expansion in an environment between average and higher density, SN explosion energy is higher than average. The spectral index is very steep 0.77 and spectrum appears to be slightly concave-up - this suggests that Cas A is a very young SNR  in which characteristic spectral forms is due to non-linear effects (see Oni{\' c} \& Uro{\v s}evi{\' c} 2015). The electron eqp magnetic field strength is 760 $\mu$G (Uro{\v s}evi{\' c} et al. 2018). According to the concept presented in this review, Cas A is a young SNR in the late free expansion phase (for higher than average environmental density, SNRs do not show the ascending  part of surface brightness evolution (P18)). This conclusion is in very good agreement with earlier confirmed facts on Cas A: it is extremely bright Galactic SNR in the free expansion phase, the so-called oxygen-rich SNR which evolves in high density medium ($\sim$ 3 cm$^{-3}$, see Arbutina \& Uro{\v s}evi{\' c} (2005), and references therein), with average magnetic field strength of $>500$ $\mu$G (Vink \& Laming 2003). The evolutionary paths from Figure 1 give slightly lower density in which Cas A expands and therefore slightly older evolutionary status.

For the youngest Galactic SNR G1.9+0.3 position in Figure 1 suggests the free expansion phase of evolution in a very rare environmental density with average SN explosion energy. The spectral index is steep $\sim$ 0.8 (Luken et al. 2020), and the electron eqp suggests magnetic field strength of 75 $\mu$G. This estimate is in good agreement with earlier confirmed facts for G1.9+0.3. We estimate here a four times lower density in which this SNR expands and due to this also an older evolutionary status. Pavlovi{\' c} (2017) presented that G1.9+0.3 is in the rising  part of its evolution. In the present period, approximately 120 years after explosion, it seems that we are witnessing approximately the fastest radio emission increase that will ever be. The rising part of evolution will be continued in the next 500 years. From Figure 1, we can conclude that G1.9+0.3 is of around maximal surface brightness, and therefore we can estimate that it is an evolutionary older SNR - not in the rising part of the free expansion evolution, but in the period of maximal brightness.

Both SNRs are very young and therefore the eqp (or constant partition) is not a valid assumption for the calculation of magnetic field value in the present moment of evolution, especially for G1.9+0.3 (see Uro{\v s}evi{\' c} et al. 2018), but after 10 kyr in future it will be. Directly from the simulations, Uro{\v s}evi{\' c} et al. (2018) obtained the magnetic field strength of 300 $\mu$G. An interesting fact to note here, for both eqp values, is that they are approximately in the order of magnitude to the correct values obtained earlier, especially for Cas A.

 Additionally, Cas A and G1.9+0.3 are not standard SNRs in order of their place on the radio surface brightness to diameter diagram. Cas A is extremely bright Galactic SNR, while G1.9+0.3 is a low brightness object given its diameter (see Figure 1). Due to this, we choose extreme objects and generally capture the reliable evolution phases for them. We miss the subphase for G1.9+0.3.

The concept presented here is based on supercomputer simulations which do not cover the radiative phases of SNR evolution. As it was stated in Raymond et al. (2020), the simple adiabatic compression of the ambient CR population, along with compression of the gas and magnetic field downstream of a radiative shock, could provide significant radio synchrotron emission. Due to this, in the radiative shocks, DSA and turbulent acceleration of CRs such as turbulent amplification of magnetic field should not be important for production of radio emission. Also, Tutone et al. (2021) suggested that reacceleration of preexisting ambient CRs provides conditions for efficient synchrotron radiation from these low velocity shocks ($v_{\rm s} < 150$ km/s). In accordance with predicted higher compression ratios on SNR shocks in the radiative phases of evolution (depend on the square of the shock Mach number), the radio spectral indices should be lower than 0.5 (for details see Oni{\' c} (2013), Uro{\v s}evi{\' c} (2014)). Additionally, old SNRs embedded in higher density medium can produce thermal bremsstrahlung radiation which can make flatter or concave-up spectra (Uro{\v s}evi{\' c} et al. 2007, Oni{\' c} et al. 2012). Moreover the high compression ratios in radiative shocks provide higher magnetic field energy densities but also higher CR energy densities. Due to
 this we can expect an approximate constant partition between the CR energy density and the magnetic field energy density during the radiative phases of evolution, until the end of evolution of an SNR (Uro{\v s}evi{\' c} et al. 2018). Bearing in mind these facts related to the radiative shocks, the analysis presented here can be extended to the entire SNR evolution.

Finally, it is reasonable to conclude that we can make a preliminary estimate of the evolutionary status of a newly observed SNR in the very simple and fast way, by using the concept presented in this review.

\section{Application of the new concept - examples}

There are several observed SNRs in the decade, mainly extragalactic, from Large and Small Magellanic Clouds (LMC and SMC), for which the new concept of determining the evolutionary status is applied. Here is a brief review of the published analyses.

The one of the first SNRs for which this concept was used is LMC SNR J0530-7007. In the study of De Horta et al. (2012), this SNR was mainly observed by Australia Telescope Compact Array (ATCA). Among other observation obtained facts, they determined the values of surface brightness and diameter for LMC SNR J0530-7007: $\Sigma_{1{\rm GHz}}=1.1\times 10^{-21}$ W/(m$^2$Hz sr), and $D=48$ pc. By using BV04 $\Sigma-D$ diagram (see Fig. 7. in De Horta et al. 2012), authors gave a suggestion that SNR J0530-7007 is in the early Sedov phase of evolution. It expands into a very low-density medium. The SN explosion energy is canonical $\sim$ $10^{51}$ erg. The eqp magnetic field strength is $\sim 50$ $\mu$G (Arbutina et al. 2012). It corresponds to relatively young to middle-aged SNR (probably in the early Sedov phase of evolution) where the interstellar magnetic field is compressed and amplified by the strong shock which expands in very low-density environment (i.e. the strength of the environment magnetic field should be lower than average). The spectral index is very steep $\alpha=0.85$, but the spectrum appears to be peaked/curved. It is in agreement with the explanation that steeper spectra correspond to young SNRs. On the other hand, this spectrum consists of only five data points, with two at the highest frequencies with probably underestimated values. It is well established that interferometers such as ATCA will suffer from missing flux at high radio frequencies due to the missing short spacings.

The next one is LMC SNR J0529-6653. The corresponding paper (Bozzetto et al. 2012a) was published at the approximately same time as the previous paper (De Horta et al. 2012). The same procedure was done. The result is similar - this SNR is in a similar evolutionary stage and looks like J0530-7007. The determined values for $\Sigma_{1{\rm GHz}}$ and $D$, again mainly from new ATCA observations, are $2.3\times 10^{-21}$ W/(m$^2$Hz sr), and $32$ pc. Again, BV04 diagram was used and inspection of the SNR location on the $\Sigma-D$ tracks leads to a possible conclusion that this SNR is in the early Sedov phase, and evolves in very low-density environment, but SN explosion energy is higher than typical: $\sim$ $2-3\times10^{51}$ erg (see Fig. 7. in Bozzetto et al. 2012a). The steep spectral index of 0.68 corresponds to an evolutionary young SNR. The eqp magnetic field calculated by using Arbutina et al. (2012) calculator is again $\sim$ 50 $\mu$G. Relatively high magnetic field strength, because of the evolution in very low-density medium (in that case MFA effects are necessary to reach 50 $\mu$G), leads to a reliable conclusion that J0529-6653 is a relatively young to middle-aged SNR, probably in the early Sedov phase.

Also, LMC SNR J0519-6902 was observed by ATCA (Bozzetto et al. 2012b). They determined $\Sigma_{1{\rm GHz}}=5.5\times 10^{-20}$ W/(m$^2$Hz sr), and $D=8.2$ pc. From the location of this SNR on BV04 plot, one more young SNR in the early Sedov phase was recognized. This object evolves in the environment of average density and the initial energy of explosion is low (see Fig. 7. in Bozzetto et al. 2012b). The eqp magnetic field value is $\sim$ 170 $\mu$G. This is the expected strength for the amplified magnetic field in a young SNR (embedded in ISM of average density). The spectral index of SNR J0519-6902 is typical for most SNRs, $\alpha=0.53$. On the other hand, the fitted line which represents spectrum is flatter because of the probably underestimated flux density at the lower frequency in the spectrum (408 MHz). Due to this with more reliable measurements at lower frequencies the spectral index should be steeper, closer to 0.6 which is typical for young SNRs. We have a reason to conclude that SNR J0519-6902 is a young SNR, in the early Sedov phase of evolution, which expands in the environment of average density.

De Horta et al. (2013) analyzed ATCA observations of Galactic SNR G308.3-1.4. This is the first Galactic SNR for which the concept given in this review was used. They calculated $\Sigma_{1{\rm GHz}}=1.1\times 10^{-21}$ W/(m$^2$Hz sr), and $D=34$ pc, by using distance (19 kpc) determined from Pavlovi{\' c} et al. (2013) $\Sigma-D$ relation. SNR G308.3-1.4 is very distant object, located on the far side of the Galaxy. Location of G308.3-1.4 on the BV04 diagram suggests that it
is in the early Sedov phase of evolution, expanding into an extremely
low-density environment with an SN explosion energy lower than the canonical SN energy of $10^{51}$ erg. The spectral index of 0.68 would be expected for a young SNR. The eqp magnetic field is $\sim$ 30 $\mu$G - the MFA effects have to be active to reach this strength because of extremely rarified IS environment in which the SNR expands and due to this very low environmental magnetic field strength. For the same reasons, we can again conclude that this is a relatively young to middle-aged SNR, probably in the early Sedov phase of evolution.

Bozzetto et al. (2013) presented a detailed study of ATCA observations of a newly discovered LMC SNR J0533-7202. From the SNR position at the BV04 diagram ($(\Sigma, D)$=$(1.1\times 10^{-21}$ W/(m$^2$Hz sr), 32.5 pc)), they suggested that SNR J0533-7202 is likely to be an SNR in the
late Sedov phase, with an explosion energy between 0.25 and $1\times 10^{51}$ erg, which evolves in an environment of density $\sim$ 1 cm$^{-3}$. The spectral index of $\sim$ 0.5, and the eqp magnetic field of 45 $\mu$G (the SNR expands into denser environment - this strength of the magnetic field can be reached exclusively by the compression on the strong shock front, the MFA does not need to be included), support an evolutionary older SNR in the late Sedov phase.

The next SNR in our analysis is LMC SNR J0508-6902. Among other observations at different wavelengths (optical and X), it was observed in radio again by ATCA (for details see Bozzetto et al. 2014a). This large SNR is in evolutionary terms similar to previous SNR J0533-7202. The large diameter is reached by evolution in the average density of environment $\sim$ 0.3 cm$^{-3}$ (BV04). Again from the BV04 diagram ($(\Sigma, D)$=$(1.4\times 10^{-21}$ W/(m$^2$Hz sr), 65.5 pc)), Bozzetto et al. (2014a) estimated that SNR J0508-6902 is likely to be an older SNR probably in transition between the Sedov and radiative phases of evolution. The spectral index of 0.6 (the steeper spectra should appear for older SNRs), and the magnetic field of $\sim$ 30 $\mu$G, again support the results obtained from the evolutionary analysis based on the $\Sigma-D$ tracks.

Bozzetto et al. (2014b) analyzed archival and own new ATCA observations for LMC SNR J0509-6731. For observed $(\Sigma, D)$=$(1.1\times 10^{-19}$ W/(m$^2$Hz sr), 7.35 pc) from BV04 plot it was estimated that this remnant is in the transition phase between the late free expansion and the early
Sedov phase, with an explosion energy of $0.25\times 10^{51}$ erg, and evolving in environment of average density of 0.3 cm$^{-3}$. This LMC SNR has similar surface brightness and diameter to Galactic Tycho and Kepler SNRs ($\Sigma_{1{\rm GHz}}=1.32\times 10^{-19}$ W/(m$^2$Hz sr), $D=9.3$ pc; $\Sigma_{1{\rm GHz}}=3.18\times 10^{-19}$ W/(m$^2$Hz sr), $D=5.2$ pc, respectively). The steep spectral index of 0.73, and the eqp magnetic field of $\sim$ 170 $\mu$G support the conclusion that this is a young SNR in transition between the late free expansion and the early Sedov phase of evolution.

The first SNR from SMC for which the evolutionary status was estimated by this concept is HFPK 334 (Crawford et al. 2014). ATCA observations gave $(\Sigma, D)$=$(3.6\times 10^{-21}$ W/(m$^2$Hz sr), 20 pc).
From BV04 $\Sigma - D$ diagram, Crawford et. al. (2014) concluded that it is a young SNR expanding in a very low density environment, with the SN explosion energy of $\sim$ $2\times 10^{51}$ erg. With the spectral index of 0.6, and the eqp magnetic field of 90 $\mu$G, all three methods support that this young SNR is in transition between the late free expansion and the early Sedov phase of evolution.

For positioning of the previously reviewed eight SNRs on the $\Sigma - D$ tracks, the BV04 evolutionary paths  were used. For the next couple of SNRs, the same concept is applied but using the new (P18) $\Sigma - D$ tracks presented in Figure 1, and upgraded eqp calculation from Uro{\v s}evi{\' c} et al. (2018).

From the radio continuum survey of the SMC by using the Australian Square Kilometre Array Pathfinder (ASCAP), two candidates for SNRs were detected for the first time (Joseph et al. 2019). The observed frequencies for the entire survey are at 960 MHz (4489 detected sources) and 1320 MHz (5954 detected sources). These two newly detected SMC SNRs are: J0057-7211 and J0106-7242. Joseph et al. (2019) applied the concept from this review to determine the evolutionary status for both newly discovered SNRs. The position of these two SNR candidates on the P18 diagram ($\Sigma = 6.38\times 10^{-22}$ and
$5.38\times 10^{-22}$ W/(m$^2$Hz sr), $D$ = 47 and 45 pc, respectively) suggests that these SNRs are in the late Sedov phase of evolution, with an explosion energy of $1-2\times 10^{51}$ erg, which evolves in a rare environment of 0.02 - 0.2 cm$^{-3}$. The spectral indices are 0.75 and 0.55, respectively. Only two observed frequencies exist and therefore these two values for spectral indices are not representative enough to justify valid conclusions. From the new eqp calculation, the magnetic field strengths are 15 and 8 $\mu$G. Finally, we can conclude that these SNRs are evolutionary older SNRs in the late Sedov phase.

Additionally, the evolutionary status of LMC SNR N103B was determined by Alsaberi et al. (2019). This SNR is probably a young type Ia SNR, similar to Galactic Tycho and Kepler SNRs, and also to LMC SNR J0509-6731 (reviewed previously in this Section). For $(\Sigma, D)$=$(6\times 10^{-19}$ W/(m$^2$Hz sr), 6.8 pc) P18 diagram suggests transition between the late free expansion and the early Sedov phase of evolution, with an explosion energy of $1-1.5\times 10^{51}$ erg, which evolves in an environment with a density of 0.02 - 0.2 cm$^{-3}$. The spectral index of N103B is 0.75, the eqp magnetic field is 235 $\mu$G. Here, it should be emphasized again that for the youngest SNRs, the equipartition assumption is not that appropriate for the determination of magnetic field strength (see Uro{\v s}evi{\' c} et al. 2018). On the other hand, the order of magnitude precision, which can be provided even for the youngest SNRs, is sufficient for the purposes of this concept. Again, all three methods suggest a young SNR in transition between the late free expansion and the early Sedov phases of evolution.

\section{Reanalysis: application of the updated $\Sigma-D$ and eqp methods}

 In Section 6, BV04 $\Sigma-D$ tracks are used for the study of the first eight SNRs. From P18 study we can use the updated evolutionary paths. The main difference between BV04 and P18 tracks is in an interesting fact obtained in P18: for SNRs with the same SN explosion energy which evolve in different ambient densities appear intersections of $\Sigma-D$ tracks for diameters between 10 pc and a few tens of parsecs, and due to this we can expect changes in the final determination of the evolutionary status for some SNRs. In BV04 diagram the crossings between tracks do not exist - their $\Sigma-D$ tracks for different densities (again for the same SN explosion energy) ends in one point - they converge with each other. Additionally, after Uro{\v s}evi{\' c} et al. (2018) study, the so-called "electron eqp", between the energy densities of CR electrons and magnetic fields should be used for the calculation of magnetic field strengths. Here we check eight previously analyzed SNRs (Section 6) in order to see whether there are any changes after using both updates.

For LMC SNR J0530-7007, the P18 diagram (Figure 1) suggests the following: expansion in a very low-density environment with canonical SN explosion energy. Instead of 50 $\mu$G, the electron eqp calculation gives 35 $\mu$G. In the very low-density environment, MFA is necessary process for reaching 35 $\mu$G. The final conclusion is the same as presented in Section 6. This is a relatively young to middle-aged SNR.

The position of LMC SNR J0529-6653in Figure 1 suggests again a very low-density environment. On the other hand, the SN explosion energy should be lower than standard one (contrary to the suggestion given in Section 6). Also at P18 $\Sigma-D$ diagram, this SNR is located close to the evolutionary path of an older SNR in the Sedov phase expanding in an ISM of average density 0.5 cm$^{-3}$, and with a higher than the average SN explosion energy $2\times10^{51}$ erg. The electron eqp magnetic field is 25 $\mu$G (instead of the previously given 50 $\mu$G). For the expansion of this SNR in an average ISM density, the magnetic field of 25 $\mu$G can be produced only by compression of the ISM magnetic field. Here, we need a third method to make a final decision - spectral slope. The spectral index is 0.68. The slope for the ordinary older SNRs in the Sedov phase of evolution should be $<0.6$. Hence, the final conclusion should be the same as given in Section 6, but this result should be taken with caution.

Now we are moving onto LMC SNR J0519-6902. Different than suggested in Section 6, from Figure 1 we can conclude that this SNR evolves in lower than average density $(0.005 - 0.02)$ cm$^{-3}$ with lower than average SN explosion energy, but in the early Sedov phase of evolution - same as concluded in Section 6. The electron eqp magnetic field is 63 $\mu$G (instead of 170 $\mu$G). Although this value is approximately 2.7 times lower than the previously obtained, for the low-density medium MFA have to be active to reach 63 $\mu$G. The final conclusion is again the same - it is a young SNR, in the early Sedov phase of evolution. The age $\sim$ 700 yr obtained in analysis of Seitenzahl et al. (2019) supports conclusion that LMC SNR J0519-6902 is young one.

The position of Galactic SNR G308.3-1.4 in Figure 1 unequivocally indicates expansion in the low-density environment $(0.005 - 0.02)$ cm$^{-3}$ with an SN explosion energy lower than the canonical SN energy of $10^{51}$ erg. The electron eqp magnetic field is 15 $\mu$G (previously determined 30 $\mu$G). In extremely rarified environment, this magnetic field strength can be provided by MFA. On the other hand, similarly to the analysis for LMC SNR J0529-6653, a possible interpretation can be evolution in a denser environment, which leads to an evolutionary older SNR in the Sedov phase. Again, the spectral index of 0.68 provides that this conclusion tentatively goes to favor young to middle-aged SNRs.

LMC SNR J0533-7202 evolves in an environment of average density (Figure 1). The SN explosion energy is slightly lower than typical. The electron eqp magnetic field is 15 $\mu$G (instead of 45 $\mu$G). This supports the same conclusion as given in Section 6 - an evolutionary older SNR in the late Sedov phase.

For LMC SNR J0508-6902, P18 diagram suggests evolution in a very low-density environment ($0.005$ cm$^{-3}$), with SN explosion energy of $2\times10^{51}$. The difference in comparison to the suggestion given in Section 6 (the evolution in an average ISM density) leads to the conclusion that this SNR is not so evolutionary old (transition between the late Sedov and radiative phases of evolution is suggested in Section 6). This SNR is in the Sedov phase of evolution with the electron eqp field of 13 $\mu$G (instead of 30 $\mu$G given in Section 6), and a spectral index of 0.6.

The position of LMC SNR J0509-6731 in Figure 1 indicates evolution in a low-density medium $\sim$ 0.02 cm$^{-3}$, with low SN explosion energy. It is again lower environmental density than suggested in Section 6. The electron eqp magnetic field strength is 95 $\mu$G (170 $\mu$G in Section 6). The final conclusion is the same - young SNR in transition between the late free expansion and the early Sedov phase of evolution. This conclusion is supported by result obtained in Seitenzahl et al. (2019) - they estimated age of $\sim$ 350 yr.

Finally, for SMC SNR HFPK 334, P18 diagram suggests evolution in the environment of average to higher density $\gtrsim$ 0.5 cm$^{-3}$, with lower SN explosion energy. The estimated ambient density in Section 6 is approximately two order of magnitude lower, and the SN explosion energy is four times higher. The electron eqp magnetic field is 38 $\mu$G (90 $\mu$G in Section 6). The spectral index is 0.6. Due to this updated analysis, this SNR is probably not a young SNR in transition between the late free expansion and the early Sedov phase of evolution, as estimated in Section 6. The presented results are consistent with the following conclusion: SMC SNR HFPK 334 is an evolutionary older SNR in the Sedov phase (with the tendency to be in transition between the late Sedov phase and radiative phases of evolution).

\section{Summary}

In this review, I presented:

\begin{description}
  \item[i)]  a brief overview of three methods related to radio observations, which can be combined for the purpose of making a preliminary estimate of the evolutionary status for an observed SNR.

  \item[ii)] explanation how to apply the new concept for the preliminary determination of the evolutionary status.

  \item[iii)] examples from literature where this concept has already been used, and additionally revision of earlier results by using the updated $\Sigma-D$ and eqp analyzes (both published in 2018).
\end{description}

At the end, I would like to emphasize the fact that this concept for the preliminary determination of the evolutionary status of SNRs is a result of approximately 20 years of work of the Belgrade SNR Research Group. The results which represent the basis for this review were published in more than 60 papers in the best astronomy and astrophysics journals. The theoretical fundamentals for all three methods, which are used in combination, were presented in: more than 25 papers for the $\Sigma-D$ analysis, more than 10 on continuum radio spectra of SNRs, more than five for eqp method applied to SNRs. The most important of them (around 20) are cited in this review. Finally, in more than 10 papers we used the concept for determining the evolutionary stages for some observed SNRs in radio, explicitly presented in this review .

%\paragraph{Paragraph headings} Use paragraph headings as needed.

%\begin{equation}
%a^2+b^2=c^2
%\end{equation}

% For one-column wide figures use
%\begin{figure}
% Use the relevant command to insert your figure file.
% For example, with the graphicx package use
%  \includegraphics{example.eps}
% figure caption is below the figure
%\caption{Please write your figure caption here}
%\label{fig:1}       % Give a unique label
%\end{figure}
%
% For two-column wide figures use

%\begin{figure*}

% Use the relevant command to insert your figure file.
% For example, with the graphicx package use

%  \includegraphics[width=0.75\textwidth]{example.eps}

% figure caption is below the figure
%\caption{Please write your figure caption here}
%\label{fig:2}       % Give a unique label
%\end{figure*}
%
% For tables use
%\begin{table}
% table caption is above the table
%\caption{Please write your table caption here}
%\label{tab:1}       % Give a unique label
% For LaTeX tables use
%\begin{tabular}{lll}
%\hline\noalign{\smallskip}
%first & second & third  \\
%\noalign{\smallskip}\hline\noalign{\smallskip}
%number & number & number \\
%number & number & number \\
%\noalign{\smallskip}\hline
%\end{tabular}
%\end{table}

\begin{acknowledgements}
I would like to thank the members of the Belgrade SNR Research Group as these are our shared results the  results set out here belong to us all: B. Arbutina, B. Vukoti{\' c}, D. Oni{\' c}, M. Vu{\v c}eti{\' c}, M. Pavlovi{\' c}, V. Zekovi{\' c}, and P. Kosti{\' c}. I also would like to thank D. Ili{\' c} and A. {\' C}iprijanovi{\' c}, our internal collaborators specialising in some other fields of astrophysics, but working with us on the emission nebulae research. The concept described in this review would never be tested on the observational data without M. Filipovi{\' c}, our main external collaborator. Also, many thanks to D. Momi{\' c}, B. Arbutina and M. Bo{\v z}i{\' c} for thought-provoking discussions and meticulous reviewing and editing of the typescript. Finally, I would like to acknowledge the referee, John Raymond, for valuable comments which significantly improved quality of this article. I am also grateful to the Ministry of Education, Science and Technological Development of Serbia for its financial support, No. of agreement 451-03-68/2022-14/200104.
\end{acknowledgements}

% BibTeX users please use one of

%\bibliographystyle{aps-bibstyle}      % American Physical Society (APS) style, author-year citations

%\bibliography{aps-nameyear.bst}                % name your BibTeX data base
%\nocite{*}

% Non-BibTeX users please use

\end{document}